\documentclass[aps,pra,twocolumn,preprintnumbers,amsmath,amssymb,superscriptaddress]{revtex4-2}
\usepackage{graphicx}
\usepackage[caption=false]{subfig}
\usepackage{epsfig}
\usepackage{dcolumn}
\usepackage{bm}
\usepackage{color}
\usepackage{float}
\usepackage[colorlinks]{hyperref}
\usepackage{verbatim}
\usepackage{algorithm}
\usepackage{algorithmic}

\hypersetup{citecolor = blue}

\def\bra#1{\left\langle#1\right|}
\def\ket#1{\left|#1\right\rangle}

\def\be{\begin{equation}}       \def\ee{\end{equation}}
\def\bea{\begin{eqnarray}}      \def\eea{\end{eqnarray}}
\def\ba{\begin{array}}
	\def\ea{\end{array}}
\def\bnum{\begin{enumerate} }
	\def\enum{\end{enumerate}}

\def\nn{\nonumber}

\def\=>{\Rightarrow}
\def\>{\rightarrow}

\def\eye2{Fathbb{I}}

\renewcommand{\>}{\rangle}

\newcommand{\al}[1]{\begin{align}#1\end{align}}
\newcommand{\eq}[2]{
	\begin{equation}
		#1 \label{#2}
	\end{equation}
}

\renewcommand{\rm}[1]{\mathrm{#1}}

\usepackage{listings}
\usepackage{color}
\definecolor{lightgray}{gray}{1}

\begin{document}
	
	\title{Variational Quantum-Neural Hybrid Error Mitigation}
	\author{Shi-Xin Zhang}
	\affiliation{Tencent Quantum Laboratory, Tencent, Shenzhen, Guangdong 518057, China}
	\affiliation{Institute for Advanced Study, Tsinghua University, Beijing 100084, China}
	\author{Zhou-Quan Wan}
	\affiliation{Institute for Advanced Study, Tsinghua University, Beijing 100084, China}
	\affiliation{Tencent Quantum Laboratory, Tencent, Shenzhen, Guangdong 518057, China}
	\author{Chang-Yu Hsieh}
	\email{kimhsieh@tencent.com}
	\affiliation{Tencent Quantum Laboratory, Tencent, Shenzhen, Guangdong 518057, China}
	\author{Hong Yao}
	\email{yaohong@tsinghua.edu.cn}
	\affiliation{Institute for Advanced Study, Tsinghua University, Beijing 100084, China}
	\author{Shengyu Zhang}
	\email{shengyzhang@tencent.com}
	\affiliation{Tencent Quantum Laboratory, Tencent, Shenzhen, Guangdong 518057, China}

	\begin{abstract}
		Quantum error mitigation (QEM) is crucial for obtaining reliable results on quantum computers by suppressing quantum noise with moderate resources. It is a key factor for successful and practical quantum algorithm implementations in the noisy intermediate scale quantum (NISQ) era. Since quantum-classical hybrid algorithms can be executed with moderate and noisy quantum resources, combining QEM with quantum-classical hybrid schemes is one of the most promising directions toward practical quantum advantages. In this work, we show how	the variational quantum-neural hybrid eigensolver (VQNHE) algorithm, which seamlessly combines the expressive power of a parameterized quantum circuit with a neural network, is inherently noise resilient with a unique QEM capacity, which is absent in vanilla variational quantum eigensolvers (VQE). We carefully analyze and elucidate the asymptotic scaling of this unique QEM capacity in VQNHE from both theoretical and experimental perspectives. Finally, we propose a variational basis transformation for the Hamiltonian to be measured under the VQNHE framework, yielding a powerful tri-optimization setup, dubbed as VQNHE++. VQNHE++ can further enhance the quantum-neural hybrid expressive power and error mitigation capacity.
		
	\end{abstract}
	
	\date{\today}
	\maketitle

	{\bf Introduction.} Variational quantum algorithms (VQA) \cite{Cerezo2020b,Bharti2021, Endo2020} are under active investigation as they require moderate quantum hardware resources and are promising candidates to deliver practical quantum advantage \cite{Arute2019, Zhong2020} in the NISQ era \cite{Preskill2018}. VQE is one of the most representative VQAs where the ground state is approximated by variational optimization \cite{Peruzzo2014, OMalley2016, McClean2016,Liu2019b,  McArdle2020, Grimsley2019, Hsieh2019} with parameterized quantum circuits.  Quantum error mitigation, as a NISQ alternative for full-fledged quantum error correction, is believed to alleviate the negative effects brought by quantum noise and deliver more reliable results for VQAs. There are already various proposals for QEM techniques  \cite{Li2017b,Temme2017, Endo2018, Kandala2019, Song2018, McArdle2019a, Chen2019a, Maciejewski2020, Bravyi2020, Barron2020, Koczor2020,Huggins2020, Huo2021,  Koczor2021} and specifically some of the proposals are based on the principle of variational optimizations \cite{Czarnik2020, Strikis2020, Cincio2020,Zlokapa2020,Lowe2020, Zhang2020b, Suchsland2020, Bultrini2021, Zhukov2021, Bennewitz2021}. However, the interplay in terms of variational optimization between VQA and QEM remains largely elusive so far. To pave the way toward more practical quantum advantages, it is natural and urgent to investigate the interplay between VQAs and QEM as well as design VQA-native QEM techniques or QEM baked-in VQAs.

\begin{figure}[t]\centering
		\includegraphics[width=0.48\textwidth]{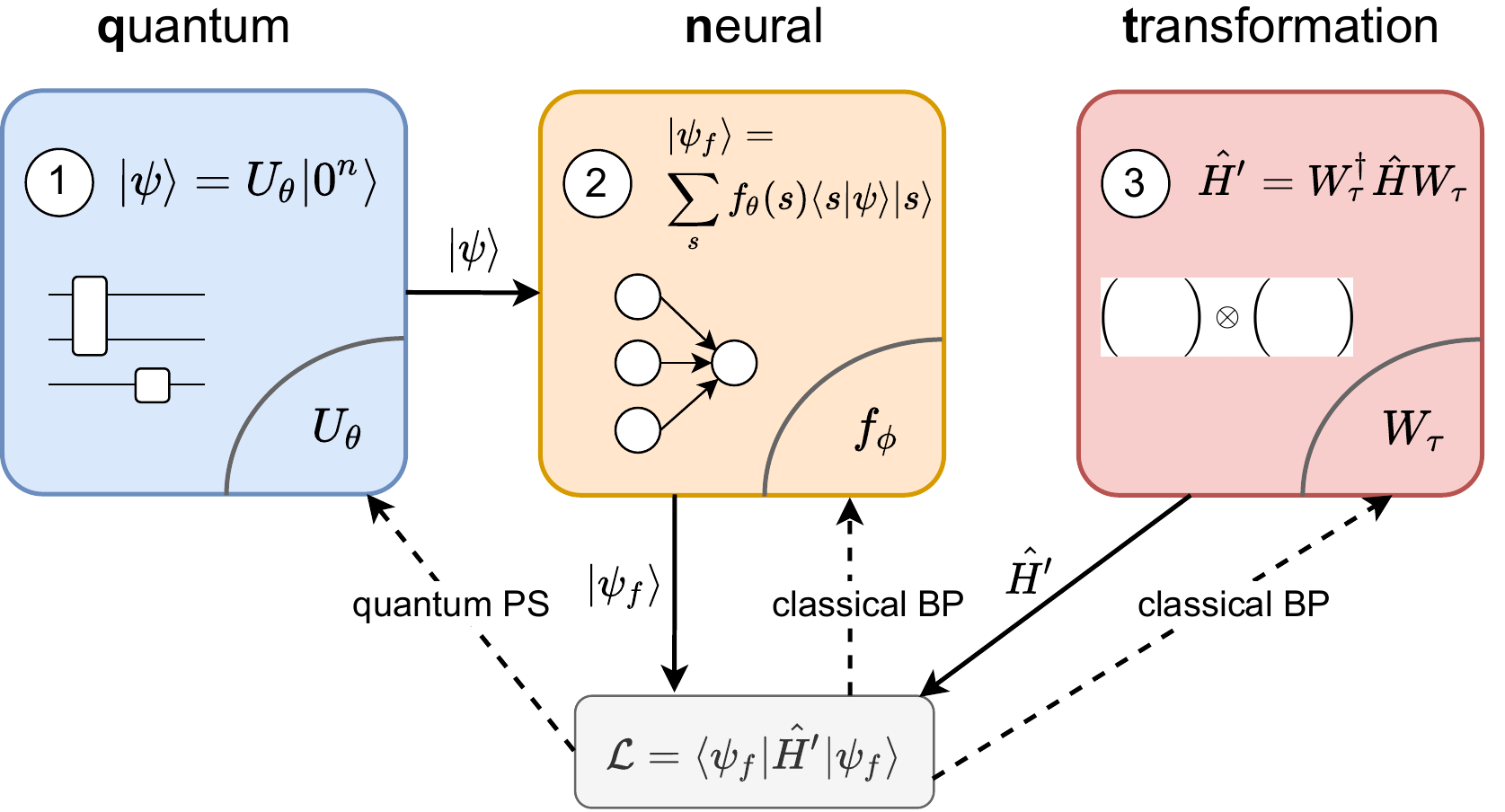}
		\caption{Schematic workflow for VQNHE++ framework where the transformed Hamiltonian approach is combined with VQNHE. The dashed lines are for gradient descent where the gradients are obtained from quantum parameter shift (PS) and classical backpropagation (BP), respectively. This tri-optimization setup enabled off-diagonal post processing for the quantum circuit output state, and thus greatly enhance the expressive power and the error resilience compared to VQNHE.}
		\label{fig:tf}
	\end{figure}

	Variational quantum-neural hybrid eigensolver (VQNHE) is a powerful VQA approach incorporating the strength of a neural network as a nonunitary post-processing module efficiently \cite{Zhang2021b}. Recently, the idea of adding a non-unitary processing module to the variational quantum eigensolver \cite{Peruzzo2014, OMalley2016, McClean2016,Liu2019b,  McArdle2020, Grimsley2019} has become popular. However, unlike all previous proposals, VQNHE not only enhances the expressive power of the VQAs but also entails just a polynomial scaling of computational resource overhead. For instance, while a previous proposal based on the Jastrow factor \cite{Jastrow1955} could enhance VQE \cite{Mazzola2019, Benfenati2021}, it requires an exponential scaling of resources overhead for the general form of Jastrow factor. In this work, we reveal another important and unique property of VQNHE: intrinsic quantum noise resilience.  Through detailed analysis, we demonstrate that the quantum noise resilience is from the introduction of the classical post-processing module and this QEM capacity is absent in the plain VQE. By utilizing the simple idea of adaptive retraining directly on noisy hardware, we obtain much more reliable energy estimations in the presence of quantum noise. In addition, by combining the transformed Hamiltonian approach in the VQNHE++ framework as shown in Fig.~\ref{fig:tf}, we further improve the expressive power and the noise resilience of the variational quantum-neural hybrid scheme, resulting in a more efficient and reliable approach for quantum simulation on noisy quantum hardware.
	
	{\bf VQNHE setup.} We first recapitulate the essence of VQNHE and then elaborate on {\it{adaptive retraining}}, a QEM protocol built on top of VQNHE in the following sections. VQNHE is an interesting example of quantum-classical hybrid schemes: it not only requires an outer classical optimizer loop but also features a classical neural network to provide the post-processing enhancement. The aim of VQNHE is the same as VQE, that is to find the ground state of a given Hamiltonian $H$ (without loss of generality, we assume H is a Pauli string below). To approximate such a ground state, we do not directly rely on the output state of a parameterized quantum circuit (PQC) $U$ as $\ket{\psi}=U\ket{0}$. Instead, we post-process the output of a PQC with a classical neural network to attain $\ket{\psi_f} = \hat{f} \ket{\psi}$. Here $\hat{f} =\sum_s f(s)\ket{s}\bra{s}$, where $f$ is a neural network with trainable weights or any general parameterized function and $s$ is a computational basis in the form of the bitstring. We note that the function $f$ can generally induce a nonunitary transformation on the quantum state. Essentially, via VQNHE, we can apply an arbitrary $2^n\times 2^n$ diagonal matrix on the output quantum state of the PQC. Previously, it was widely believed that the experimental implementation for accurate estimations on the energy  $\bra{\psi_f}\hat{H}\ket{\psi_f}/\bra{\psi_f}\psi_f\rangle$ requires exponential time. However, as explicated in Ref \cite{Zhang2021b}, this energy estimation can be accurately and efficiently obtained with only a polynomial scaling of hardware resources. 
	
	We now describe the experimental protocol for measuring the Hamiltonian expectation with the classical post-processing scheme  $f$ on the output of PQC $U$. 
	Without loss of generality, we only show how to measure the expectation for a Pauli string $H$, as the expectation for a general Hamiltonian can be decomposed into a weighted sum of a polynomial number of different Pauli strings in most realistic cases. We define the expectation value for the Pauli string $H$ with classical post-processing $f$ as:
	\eq{	\langle \hat{H}\rangle_{\psi_f} = \frac{\bra{\psi_f}\hat{H}\ket{\psi_f}} {\bra{\psi_f}\psi_f\rangle}. }{}
	
	The Pauli string is expressed as $H=\prod_{k=1}^n H_k$, where $H_k$ correspond to local Pauli operator I, X, Y, or Z. We denoted the set $i/Z = \{i\vert H_i=Z\}$, namely, the qubit indices where $H$ hosts $Z$ operator.
	
	If the Pauli string contains no X or Y operator, the energy estimation is straightforward and is given as: 
	\eq{	\langle \hat{H}\rangle_{\psi_f} = \frac{\sum_{s\in U} f(s)^2 \prod_{i / Z} (1-2s_i)}{\sum_{s\in U} f(s)^2},}{}
	where $s\in U$ denotes the results collected on the computational basis of circuit $U$, i.e. $s$ is the measurement bitstring results for $U$ circuit. 
	
If the Pauli string contains X or Y operator, we call the first qubit that hosts X or Y operator in the Pauli string as the sign qubit and relabel the qubit as qubit 0 below for notation convenience. We build a measurement circuit block $V$ which is attached after the PQC $U$. The building rule for $V$ is: (1) We apply a control-X gate with control on the sign qubit and target for each qubit $i/X$. We also apply a control-Y gate with control on the sign qubit and target for each qubit $i/Y$. (2) We measure the sign qubit in X or Y direction depending on the operator type on the sign qubit. In other words, we apply an H gate or $e^{-i\pi/4 X}$ gate on sign qubit in the circuit $V$. The energy can be estimated in an unbiased and efficient manner as:	
	\eq{
		\langle \hat{H}\rangle_{\psi_f} = \frac{\sum_{s\in UV}{(1-2s_0)\prod_{i/Z}(1-2s_i) f(0s_{1:n-1})f(1\widetilde{s_{1:n-1}})) }}{\sum_{s\in U}{f(s)^2}},
	}{eq:main}
	where the bitstring $s$ in the denominator is drawn from the PQC $U$ and bitstring $s$ in the numerator is drawn from the PQC with the measurement circuit $V$ appended. $\tilde{s}$ is for bitstring with bit-flip on $s$ on qubit indices $i/X$ and $i/Y$. ($\vert \tilde{s}\rangle \propto H\vert s\rangle$). $1\widetilde{s_{1:n-1}}$ implies that for each bitstring $s$ collected from the experiments, we set the first bit as $1$ and flip the following bits if the Pauli string has an X or Y operator on the corresponding position.
	
	In the above, we introduce the scalable protocol on expectation evaluation in VQNHE. To train the model, we need to evaluate the gradient for both neural network and variational circuit parameters if the gradient-based optimizer is adopted. In terms of the circuit parameters, the conventional parameter shift rule still applies since we can regard the process as a plain VQE with the Hamiltonian to be evaluated as $f^\dagger H f /\langle \psi\vert f^\dagger f\vert \psi\rangle$. When the measurement results in the form of a collection of bitstrings are fixed, the energy evaluation function as indicated by Eq \eqref{eq:main} is a purely classical function with neural parameters, whose gradients can be evaluated via automatic differentiation (back-propagation) method numerically.
	
	In summary, VQNHE jointly optimizes the parameters in the PQC $U$ and the classical post-processing module $f$. As an approach combining the advantages from both VQE and neural variational Monte Carlo (VMC) \cite{Carleo2017,Deng2017a, Carleo2018, Cai2018a, Pfau2019a,Hermann2019a,Zhang2019b}, this new setup offers a state-of-the-art approximation on the ground state energy for various quantum spin systems and quantum molecules with a provable bound on the efficiency for the computational complexity \cite{Zhang2021b}.

	{\bf Retraining energy gain as a measure for QEM capacity.} We investigate the VQNHE performance on both noisy quantum simulators and real quantum hardware. The quantum noise deteriorates the accuracy of the energy estimation and thus compromises the superior performance that could be attained in an ideal situation, such as a noise-free simulation. Interestingly, we find that VQNHE exhibits inherent noise resilience to a certain extent. Namely, when training the VQNHE in a noisy environment, the neural network can adjust its weights, implicitly mitigating noise-induced disruptions. We term the optimization on noisy hardware the {\it adaptive retraining}. The QEM capacity of VQNHE can be measured by the difference of energy estimations $\delta E = E_{\phi} - E_{\phi_0}$, where $E_\phi$ is the energy estimation with neural weights $\phi$, and $\phi(\phi_0)$ is the optimized weights with(out) the presence of quantum noise. Apart from the neural retraining, we can also investigate the energy gain when retraining the PQC or jointly retraining the PQC and the neural network. The energy gains are defined as $\delta E = E_{\theta}-E_{\theta_0}$ and $\delta E = E_{\theta, \phi}-E_{\theta_0,\phi_0}$, respectively, where $\theta(\theta_0)$ is the set of optimized parameters in the PQC trained with(out) noise and both energies are all evaluated in the noisy setting. The more negative energy gain indicates better noise resilience as it reflects the amount of energy that is further lowered via retraining from ideal parameters in noisy devices. 
	
	It is worth noting that the so-called retraining can start from any weight initialization, especially in the joint retraining case. The noiseless optimal parameters are only used to define the retraining energy gain theoretically, and are not necessary for the practical QEM. We use this metric instead of the absolute energy to characterize the QEM capacity since we need to separate the contribution of error mitigation from the expressive power enhancement. As we discussed below, the energy gain metric also shows some nice scaling behaviors which can be explained from theoretical understanding.  
	
\begin{figure}[t]\centering
		\includegraphics[width=0.47\textwidth]{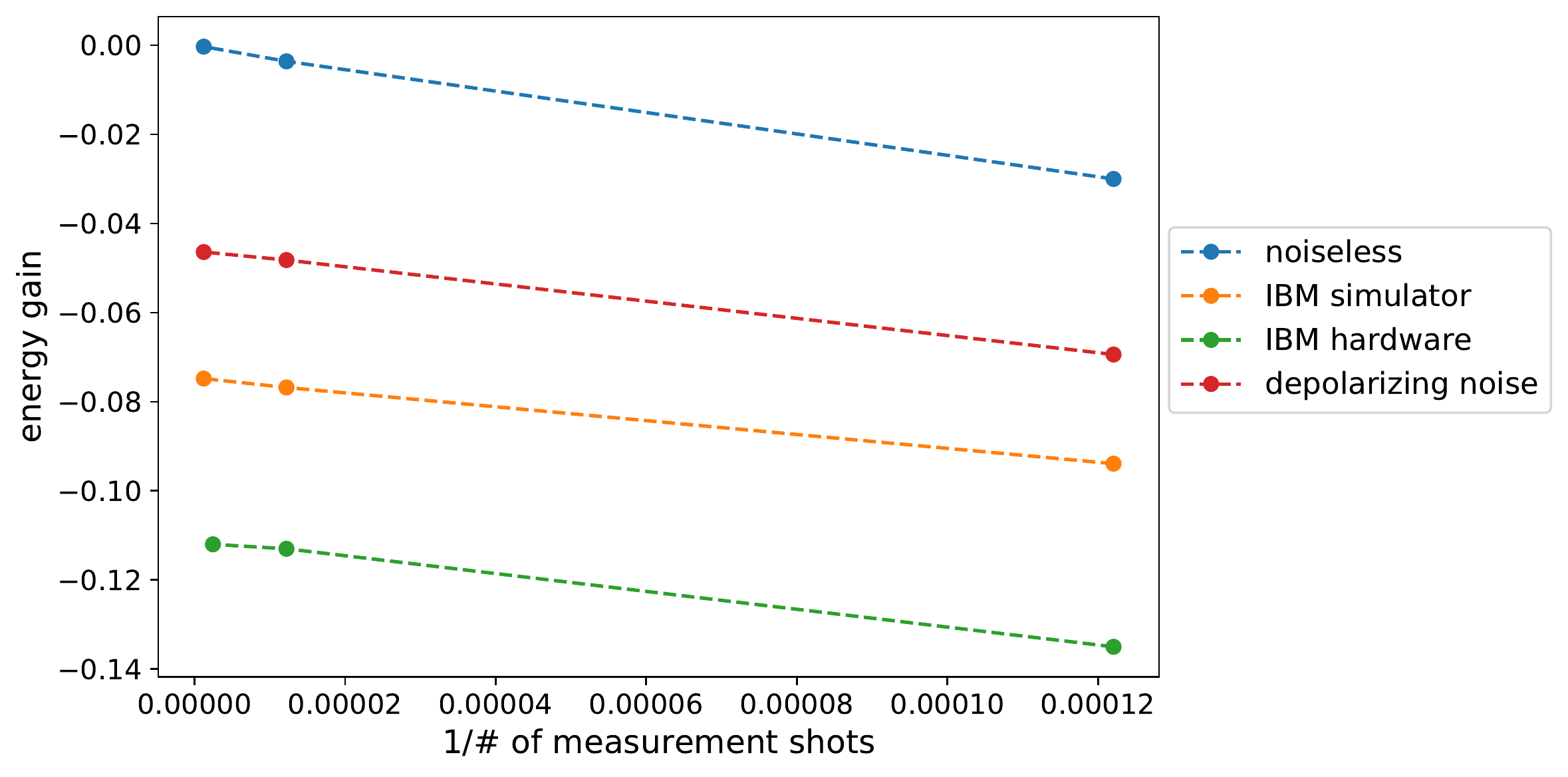}
		\caption{Scaling between biased retraining energy gain and the number of measurement shots. The system under investigation is a five-qubit TFIM and the test environments include the noiseless case, artificial noise case, and real noise cases from the IBM simulator and real quantum hardware. The biased energy gain scales linearly with the inverse of the number of measurement shots. And the intercept corresponds to the intrinsic QEM capacity in each case.}
		\label{fig:biasscaling}
	\end{figure}

	{\bf Biased retraining on the classical module.}
	In real experiments, the energy is estimated from a collection of measured bitstrings with finite sampling errors. The number of measurement shots required is often large, especially when near optimum or due to the vanishing gradients \cite{McClean2018, Wang2020, Cerezo2020a}. Therefore, it is expensive to run full unbiased retraining. To this end, we propose a very cost-efficient alternative, i.e. biased retraining, which only retrains the neural network part. In the biased retraining, instead of executing the PQC at each training epoch, we fix the bitstring measurement results during the retraining. Since the bitstring results are fixed (with just a finite number of shots), they are biased with measurement uncertainty. As a result, the obtained biased retraining energy gain has two components: the intrinsic QEM and overfitting to the biased measurements. With IBM device-compatible noisy simulations and real IBM hardware experiments, we demonstrate that the average QEM capacity scaling for the biased neural retraining is $\overline{\delta E}\propto B+A/M$, where $M$ is the number of fixed measurement shots and the constant $B$ stands for intrinsic QEM capacity. The intrinsic QEM part remains when the number of measurement shots is taken to infinity  $M\rightarrow\infty$ where the training bias induced by the finite shot noise vanishes. Fig.~\ref{fig:biasscaling} shows the energy gain results from biased retraining under different noise models including the real hardware experiment, and all results conform to the scaling relation. (See SM Sec.~\ref{smsec:biased} for the data and experiment setup details for the Hamiltonian, circuit ansatz, neural network structures, etc.)

	{\bf QEM capacity scaling with the noise strength.}
	To investigate the energy gain with a tunable noise strength, we utilize a simple depolarizing error model, where an isotropic depolarization of strength $p$ is attached after each two-qubit gate. The one-dimensional five-site transverse field Ising model (TFIM) with an open boundary condition is then utilized as the VQNHE target Hamiltonian: $\hat{H}=\sum_{i=1}^{n-1} Z_iZ_{i+1}-\sum_{i=1}^n X_i$. And the numerical simulation is implemented using TensorCircuit \cite{Zhang2022}. We study the scaling relation between the energy gain due to the intrinsic QEM and the effective overall noise strength $p_{\text{eff}}$ of the depolarization. The overall depolarizing probability $p_{\text{eff}}$ is measured by the energy ratio from the PQC output:  $1-p_{\text{eff}}=E_{\text{noise}}/E_{\text{noiseless}}$.
	
	Firstly, retraining solely on the quantum part cannot improve the final energy estimation. With only quantum retraining, the optimal energy estimation in the noisy case is always $(1-p_{\text{eff}})E_{\text{noiseless}}$. This fact implies that the adaptive retraining QEM procedure is unique to the pipelines with classical post-processing and plain VQE is not quantum noise resilient in the sense of adaptive retraining.
	
	For depolarizing strength $p=0.005, 0.01, 0.015, 0.02$, the effective overall error strength is correspondingly $p_{\text{eff}}=0.017, 0.033, 0.049, 0.065$ with our circuit ansatz with optimal circuit weights and the intrinsic QEM energy gain from neural retraining are $\delta E=-0.0031, -0.012, -0.026, -0.044$, respectively. The scaling relation for the QEM energy gain with neural retraining is thus given by $\delta E \propto p_{\text {eff}}^2$.
	
	\begin{figure}[t]\centering
		\includegraphics[width=0.42\textwidth]{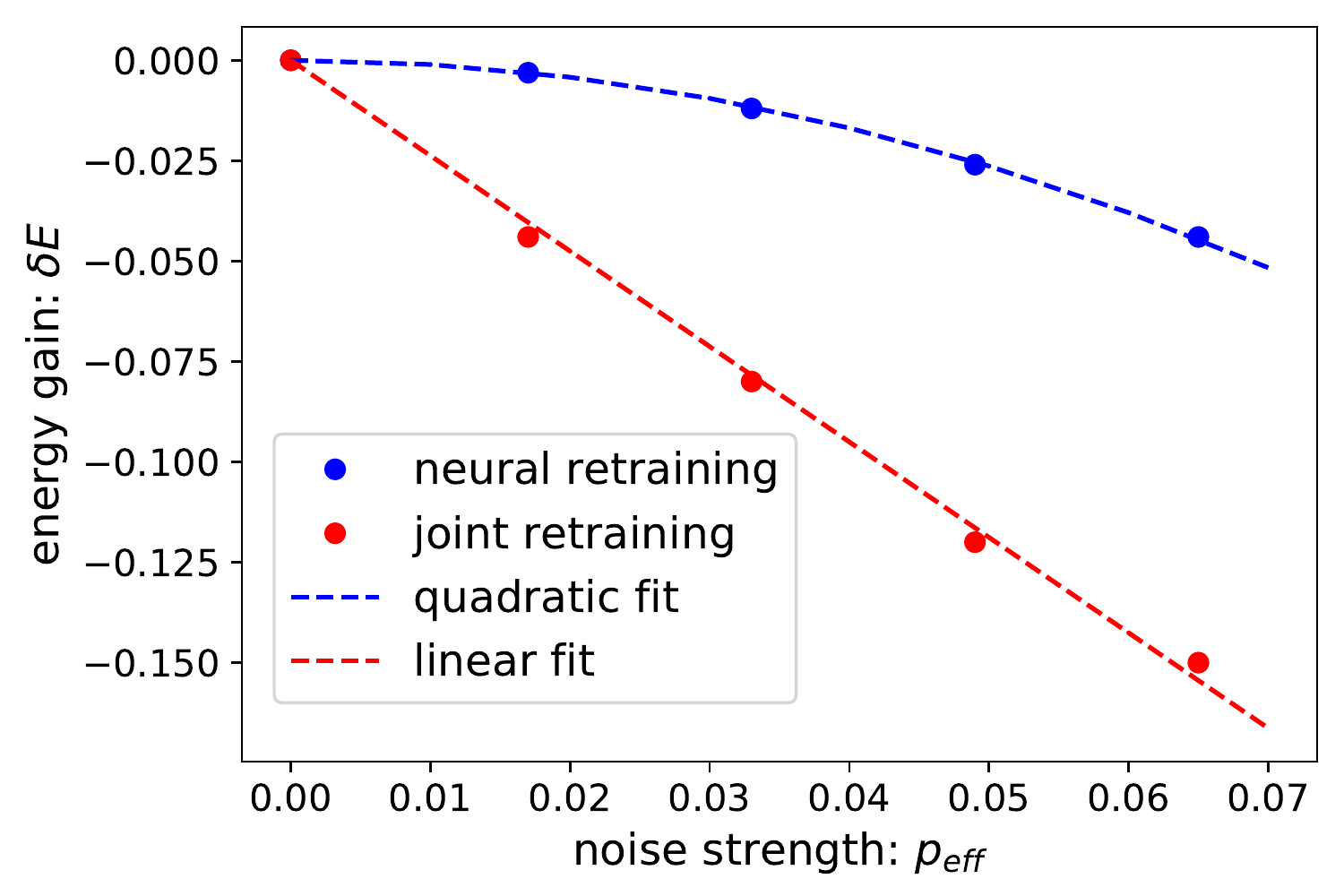}
		\caption{Scaling between retraining energy gain and the noise strength. The system is five-qubit TFIM and the underlying error model is depolarizing noise. The QEM capacity due to neural retraining is quadratic with the noise strength and is thus much weaker than joint retraining, which has linear scaling against the noise strength.}
		\label{fig:scaling}
	\end{figure}

	The intrinsic QEM capacity with neural retraining scales quadratically with $p_{\text{eff}}$, which is the reason behind the biased neural retraining scaling with the number of measurements we observed before. Note that the effective noise strength $p_{\text{eff}}$ can only be approximately estimated in experiments from finite measurement shots. We take $p_{\text {eff}}$ as a random variable and the energy gain is $\delta E\propto \langle p_{\text {eff}}^2\rangle= \langle p_{\text{eff}}\rangle^2+ \Delta p_{\text {eff}}$, where $\Delta p_{\text {eff}} \propto 1/M$ is the square deviation of the estimation on $p_{\text {eff}}$ due to finite measurement shots.  Therefore, the energy gain after retraining follows a simple scaling form of $A/M+B$.
	
	In addition, we evaluate the QEM energy gain from joint retraining, with $\delta E = -0.044$,$ -0.080$, $-0.12$, $-0.15$ for effective error strengths listed above. Therefore, the scaling of QEM capacity with the quantum noise strength from joint retraining is linear instead of quadratic: $\delta E \propto p_{\text{eff}}$. The different scaling forms are sketched in Fig.~\ref{fig:scaling} for both neural retraining and joint retraining. (See more rigorous fitting in log-log scale in SM Sec. \ref{smsec:logscale}).

	{\bf General picture for the QEM scaling.}
	To understand the above QEM scaling relations, we first investigate a minimal model involving just one qubit which permits direct analytical analysis. The results are consistent with the observed scaling relation (see SM Sec.~\ref{smsec:onequbit} for details).
	
	We now discuss the theoretical mechanism behind different QEM capacity scalings.
	Suppose that the ideal output of a PQC is the exact ground state as $\rho_0=\vert \psi_0\rangle\langle \psi_0\vert$. And the mixed state from the PQC in the presence of depolarizing noise of strength $p$ is $\rho$. % (we still consider the depolarizing channel).
	The post-processing module is a nonunitary transformation $\hat{f}$ with non-zero elements only appearing in the diagonal. % (only real-valued $f$ considered below).
	The energy gain with neural retraining is thus $\delta E =E_{QEM}-E_{N}$. Here $E_N= (1-p)E_0$ where $E_0 = \rm{Tr}(\rho_0 H)$ is for the exact ground state.
	%\eq{
	%\delta E =E_{QEM}-E_{N} =\min_{\phi} \frac{Tr(\hat{f}_\phi\rho\hat{f}_\phi\hat{H})}{Tr(\hat{f}_\phi\rho\hat{f}_\phi)}-Tr(\rho\hat{H})
	%}{}
	%We consider the depolarizing noise $\rho=(1-p)\rho_0+p/2^n I$ where $\rho_0 = \vert \psi_0\rangle\langle \psi_0\vert$ is the pure ground state.
	We expand the energy terms as $E_{QEM} = E_{QEM}^{(0)} + E_{QEM}^{(1)}p + E_{QEM}^{(2)}p^2+\cdots$ and keep up to the first order of $p$, namely, as long as we have shown that the zeroth and first order of $p$ in the energy gain is zero, the energy gain scaling is at most $p^2$.

	Under depolarizing channel $p$, we have:
	\eq{
		E_{QEM} = \frac{\rm{Tr}(\hat{f}\rho_0\hat{f} \hat{H})(1-p)+\rm{Tr}(\hat{f}\hat{H}\hat{f})p/2^n}{p+\rm{Tr}(\hat{f}\rho_0\hat{f})(1-p)}.
	}{}

	The optimized neural module $f=I$ when $p=0$. We assume the optimized $f\approx I +p f_1$ to the first order, where $f_1$ is a constant matrix.
	
	The zeroth order of $p$ in the energy gain is trivially zero: $E_{QEM}^{(0)}=E_0$. %and thus $\delta E^{(0)} = 0$.
	Now consider the first order of $p$, and we have $E_N^{(1)}=-E_0$ and
	\al{
		E_{QEM}^{(1)}=\rm{Tr}(f_1\rho_0H)-E_0 \rm{Tr}(f_1\rho_0) \nn\\ +\rm{Tr}(\rho_0f_1H)-E_0\rm{Tr}(\rho_0f_1)-E_0.
	}
	Note that
	\eq{\rm{Tr}(f_1\rho_0 H)-E_0\rm{Tr}(f_1\rho_0) = \langle \psi_0\vert Hf_1\vert \psi_0\rangle- E_0\langle \psi_0\vert f_1\vert \psi_0\rangle =0}{}
	thus we have $E^{(1)}_{QEM}=-E_0$, independent of $f_1$. Therefore, the first order energy gain vanishes $\delta E^{(1)}=0$ as well, indicating that the energy gain scales at most quadratically with the noise strength $p$.
	To summarize, the first order of $p$ in the energy gain is canceled as long as the retraining trajectory can be understood from a simple perturbation, i.e. the optimized post-processing module $f$ is analytically connected to the ideal one $I$ as the noise $p\rightarrow 0$.
	
	To explain why a linear gain emerges in the joint retraining, we note that the perturbative picture fails under the joint-training scenario. As long as we allow joint training, there are infinitely many optimal solutions, constituted by appropriate combinations of PQC and neural-network weights to essentially yield the same output state in the noiseless case. We are no longer restricted to the unique solution as in the neural retraining case, where the PQC generates the exact ground state with an identity neural network (NN). Instead, even when the PQC generates other quantum states than the true ground state, an appropriate post-processing neural module $f$ can still post-process to the ground truth. Therefore, in the ideal case $p=0$, we have infinitely many combinations of PQC states and neural solutions that would collaboratively lead to the correct ground state energy. 
	When noise $p>0$ is introduced, the responses to quantum noise are different and the energy degeneracy (of many possible combinations of PQC and NN setups) is broken. %This process is similar to the famous symmetry breaking picture.
	Therefore, the optimal solution in the weak noise case is not connected to the identity one $f=I$ in the joint retraining case. 
	In other words, the optimized $f$ cannot be simply described by $f=I+pf_1$ where the derivation based on the perturbation picture fails and the first order energy gain emerges.
	
	In summary, the energy gains in neural retraining and joint retraining come from different sources and they can be understood using clear and unified physical pictures. The neural retraining perturbatively improves the noisy energy estimation by smoothly shifting the classical module $f$ away from identity $I$. On the contrary, the joint retraining improves energy estimation by breaking the degeneracy of infinitely many possible combinations of PQC and neural setups and selecting the most error-resilient one.

	{\bf VQNHE++: Tri-optimization with parameterized transformed Hamiltonian.}
	VQNHE is a bi-optimization setup, where both parameters $\theta$ in the PQC and parameters $\phi$ in the neural network need to be optimized. The post-processing function $f$ can greatly alter the output states by the PQC. However, 
	$\hat{f}$ is effectively a diagonal matrix, which certainly cannot represent a universal quantum operation. Therefore, such retraining of the neural post-processing module can only partially mitigate the quantum noise effects. Since a universal nonunitary quantum operation is NP hard to implement in terms of quantum resources, we instead introduce a parameterized gauge Hamiltonian approach to enhance the mitigating power of the post-processing quantum channel $\hat{f}$.
	
	Suppose that $\hat{W}$ is a unitary transformation, then the transformed Hamiltonian $\hat{H'}=\hat{W}^\dagger \hat{H}\hat{W}$ shares the identical spectrum with $\hat{H}$, and thus the ground state energy is the same as that of $\hat{H}$. Therefore, we can utilize VQNHE to simulate the ground state of the transformed Hamiltonian by identifying the Pauli strings in the newly transformed Hamiltonian as observables. To efficiently implement this idea, we require that the number of Pauli strings in $\hat{H'}$ scales polynomially with the system size $n$. This requirement restricts the possible forms of gauge transformation for $\hat{W}$. For local Hamiltonian such as quantum spin models, $\hat{W}$ can be in the form of single-qubit rotation gates $\hat{W}=\prod_i \exp({i\tau_i P_i})$, where $P_i$ is the Pauli gate X, Y or Z. Some special forms or structures of parameterized tensor networks can also play the role of the Hamiltonian transformation with better expressiveness and controllable overhead. (See more details on gauge transformation design and analysis in SM Sec. \ref{smsec:gauge}.)
	
	The experimental protocol for VQNHE++ is a straightforward combination of the protocol for VQNHE as explained before and the experimental protocol for the transformed Hamiltonian approach. Namely, we classically track the parameterized transformed Hamiltonian since there are only polynomial Pauli string terms after the transformation which can be efficiently obtained and manipulated on classical computers via simple Pauli matrix commutation algebra.  We then can apply VQNHE framework on the transformed Hamiltonian instead of the original Hamiltonian (e.g. Eq \eqref{eq:eg} for a TFIM transformed Hamiltonian). Similarly, with fixed measurement results, the energy evaluation forward pass is a pure classical function with neural weights and transformation parameters. Therefore, the gradients for both types of parameters can be obtained efficiently by automatic differentiation.  From a higher level perspective, we apply the diagonal post-processing on the output quantum state from the PQC using VQNHE protocol as a Schr\"odinger picture operation and apply the parameterized gauged transformation on the target Hamiltonian as a Heisenberg picture operation. VQNHE++ framework is scalable as it still maintains the polynomial computational complexity.

	\begin{figure}[t]\centering
		\includegraphics[width=0.45\textwidth]{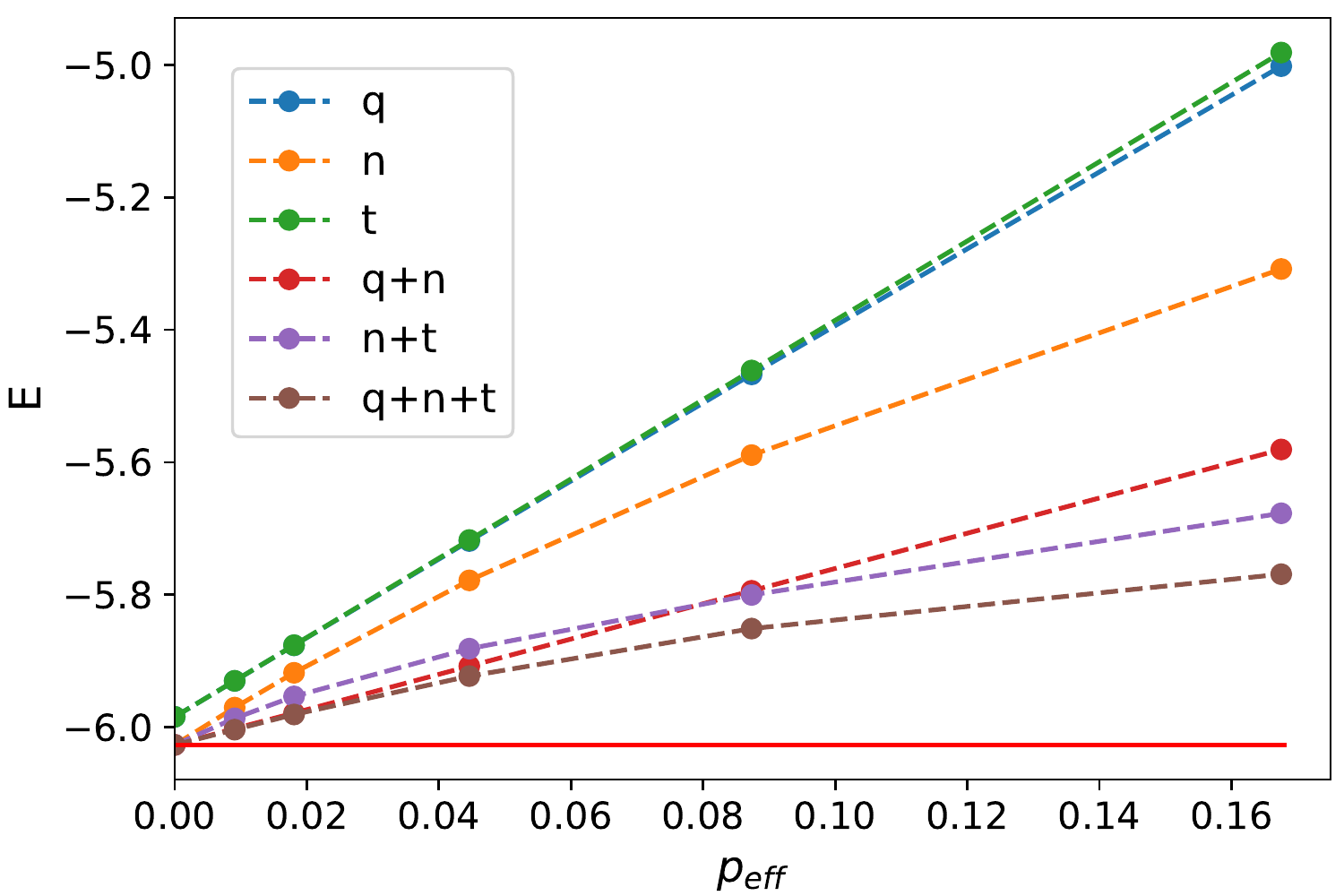}
		\caption{VQNHE++ results for TFIM on the noisy circuit with depolarizing noise model using different adaptive retraining strategies. The depolarizing noise strength is characterized by $p_{\text{eff}}=1-E_n/E_0$, where $E_n$ ($E_0$) is energy estimation with(out) noise. Here q is for retraining on the PQC part, n is for retraining on the neural network part and t is for retraining on the parameterized gauge transformation part. We omit the result of no retraining and q+t as they are both very similar to retraining PQC (q). The solid red line indicates the exact ground state energy.}
		\label{fig:sixopt}
	\end{figure}
	
	From a theoretical perspective, we now have the energy estimation as:
	\al{\langle \hat{H}\rangle &= \rm{Tr}\left(\hat{f}_\phi\rho_\theta \hat{f}^\dagger_\phi (\hat{W_\tau}^\dagger \hat{H} \hat{W}_\tau)  \right)/\rm{Tr}\left(\hat{f}_\phi\rho_\theta \hat{f}^\dagger_\phi  \right)\nn\\&=\rm{Tr}\left((\hat{W}_\tau \hat{f}_\phi)\rho_\theta (\hat{W_\tau}\hat{f}_\phi )^\dagger \hat{H}  \right)/\rm{Tr}\left(\hat{f}_\phi\rho_\theta \hat{f}^\dagger_\phi  \right).
	}
	%where $\rho$ is the density matrix of the PQC output on the noisy hardware.
	Therefore, the transformed Hamiltonian setup with VQNHE essentially gives a more powerful variational post-processing channel than the plain diagonal matrix $\hat{f}$. The new effective post-processing operation is $\hat{W}_\tau \hat{f}_\phi$, which has non-vanishing off-diagonal contributions. The enhanced post-processing capacity implies better performance on ground state energy optimization and quantum error mitigation as the freedom in the new ansatz is strictly larger than VQNHE. An intuitive limit is by considering $\hat{W}$ as a diagonal transformation for the original Hamiltonian $H$, i.e., the transformed Hamiltonian $\hat{H'}$ is a diagonal matrix. In that case, we can train a diagonal $f$ to successfully project any PQC output $\rho$ to the exact ground state and thus free from any quantum noise. 
	%The only restriction comes from the practical consideration that the gauge transformation $\hat{W}$ must keep polynomial number of Pauli string terms overhead in the transformation.
	
	In the plain VQE, the transformed Hamiltonian operation can be directly implemented on the circuit instead of tracking the new transformed Hamiltonian virtually (say for local Ry transformation, we directly apply one layer of parameterized Ry gate at the end of the PQC). Nevertheless, there is still a subtle difference between the direct implementation of the transformation on the circuit (Schr\"odinger picture) and the transformed Hamiltonian tracked classically (Heisenberg picture): the latter is free from quantum noise for the transformation part. In the VQNHE setup, such parameterized transformation cannot be implemented on the circuit as there is an uncommutable neural diagonal module in between. The operation order in VQNHE++ is: PQC $U_\theta$ + neural post-processing $f_\phi$ + gauge transformation $\hat{W}_\tau$. If we naively implement the gauge transformation on the circuit, the order is instead PQC $U_\theta$ + gauge transformation $W_\tau$ + neural post-processing $f_\phi$. The two orders give totally different effective operations even without quantum noise as they are uncommuting with each other, and the latter is more trivial as the transformation can be absorbed into the circuit in the noiseless limit.
	
	We consider the five-qubit TFIM as a specific example to demonstrate the workflow and illustrate the benefits of the transformed Hamiltonian approach. The model is aligned with the one we utilized in VQNHE experiments and is compatible with public IBM hardware devices. We take the gauge transformation $\hat{W}=\prod_i \exp{i\tau_i Y_i}$, and the corresponding transformed Hamiltonian is
	\al{
		\hat{H'_\tau}= &\sum_i ( \cos 2\tau_i\cos 2\tau_{i+1} Z_iZ_{i+1} +\sin2\tau_i\sin 2\tau_{i+1} X_iX_{i+1}\nn\\-&\sin 2\tau_i\cos2\tau_{i+1} X_i Z_{i+1} -\cos 2\tau_i \sin 2\tau_{i+1} Z_i X_{i+1}\nn\\ -&\cos 2\tau_i X_i-\sin 2\tau_i Z_i),\label{eq:eg}
	}
	which contains a polynomial number of Pauli string terms.
	We utilize the PQC ansatz of a layered form [H, ZZ($\theta_1$), Rx($\theta_2$), XX($\theta_3$), Ry($\theta_4$)] (See SM Sec.~\ref{smsec:notation} for circuit ansatz representation notation).

	With the introduction of the transformed Hamiltonian on top of the VQNHE setup, we are now equipped with more options for noisy adaptive retraining.  We run adaptive retraining for all combinations of {\bf q}uantum module, {\bf n}eural module and {\bf t}ransformation module. The absolute energy estimated after each kind of retraining is displayed in Fig.~\ref{fig:sixopt}. It is worth noting that the line for the quantum-only retraining also nearly coincides with the line of no retraining and retraining on quantum and transformation parts (not shown), since the quantum part itself cannot be tuned to minimize the depolarizing error as we mentioned before, and the transformation part can be absorbed into the last layer of Ry in the PQC trivially when the neural module is fixed to identity.
	
\begin{figure}[t]\centering
		\includegraphics[width=0.46\textwidth]{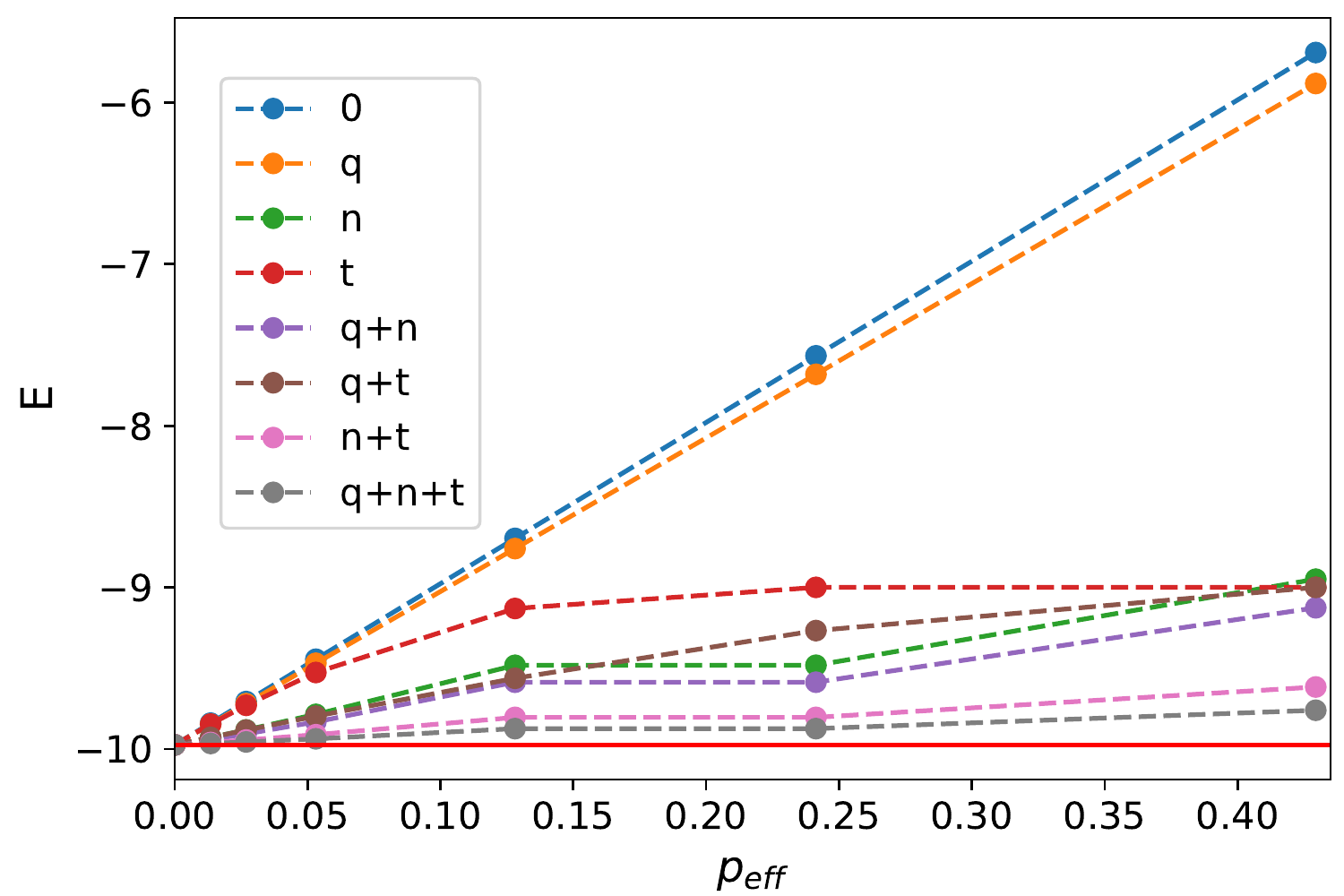}
		\caption{VQNHE++ results for 1D Heisenberg model with overall depolarizing noise $p_{\text {eff}}$. 0 indicates the energy estimation with noiseless optimal weights (no retraining). q, n, t is for retraining on the PQC, neural network and parameterized transformations, respectively. The solid line is the exact ground state energy for the simulated system.}
		\label{fig:hei}
	\end{figure}

	The most crucial insights from Fig.~\ref{fig:sixopt} are that the n+t retraining delivers a similar QEM performance as the joint retraining (q+n), and that the QEM capacity for the n+t retraining is even stronger than the conventional joint retraining (q+n) when the overall noise strength is high. Since the PQC is fixed in the n+t retraining scheme, we can carry out the fast biased retraining based on the same set of measurement results from the PQC, similar to the biased neural retraining (n) case we discussed before.
	Therefore, the classically tractable biased retraining combining the neural post-processing and parameterized Hamiltonian transformation can achieve competitive QEM results as joint retraining but avoid issues such as finite sampling errors or quantum gradient vanishing (barren plateau issue). We also report the transformed Hamiltonian approach on the Heisenberg model with various quantum noise models and obtain better error mitigation results. Fig.~\ref{fig:hei} shows the results of different retraining strategies for Heisenberg model VQNHE. (See SM Sec. \ref{smsec:heisenberg} for setup details and other results.) The mitigated energy is at least $E=-9$ in this case since even for the fully mixed state $\rho=I/2^n$, the transformation, as a quantum channel effectively, can project the system to an averaged energy of $-9$. The consistently promising results for different Hamiltonian systems and under different noise models demonstrate the universal capability of VQNHE++ for ground state problems with built-in error mitigation power.
	
	{\bf Discussions:} In this work, we mainly focus on the noise resilience aspect of VQNHE. The proposed error mitigation method integrated with VQNHE is very promising as it requires fewer hardware resources compared to other well-established QEM schemes. The advantage of resource efficiency is especially prominent for the biased retraining, which only requires the same amount of measurement shots and hardware resources as one round of energy estimation. On the contrary, zero noise extrapolation (ZNE) \cite{Li2017b,Temme2017} , one of the most common QEM techniques,  needs to be conducted on the hardware of different noise strengths. %Besides, if we fix the PQC weights, then the zero noise extrapolation limit is biased since the PQC is suboptimal in the noiseless limit. 
	Moreover, virtual distillation method \cite{Koczor2020,Huggins2020, Koczor2021, Huo2021}, which prepares multiple copies of the state, and quasi-probability method \cite{Temme2017, Strikis2020}, which requires tomography on the gates and the exponential scaling with the sampling positions, take much more hardware resources and running times. The neural error mitigation scheme proposed in Ref \cite{Bennewitz2021} essentially provides a good initialization for the neural VMC with expensive resource requirements for the state tomography. And the second stage of the scheme  in Ref \cite{Bennewitz2021} is a purely classical VMC training with no input from the PQC. On the contrary, the noisy PQC is always one part of the quantum state generation pipeline in our case. 
	
   Several further comments are in order. Firstly, our QEM proposal is strongly correlated with the VQNHE setup and rooted in the energy variational principle.
	Therefore, the current proposal is not a universal error mitigation method for universal quantum computing tasks. 
	Secondly, the current QEM scheme can be easily combined with other common error mitigation techniques for further error reduction. Since most QEM schemes focus on error mitigation for the expectation values (from PQC) of some observables, they are compatible with the adaptive retraining for VQNHE. Specifically, we have successfully combined a technique of readout error mitigation with the retraining scheme (see the results in SM Sec. \ref{smsec:biased}). 
	Thirdly, it is also interesting to observe how robust amplitude estimation \cite{Wang2021a} with parameterized likelihood can also exhibit error mitigation capability \cite{Katabarwa2021}. There are some similarities between VQNHE and robust amplitude estimation conceptually. Both methods are amplitude amplification algorithms where the classical neural module acts as the amplifier in VQNHE as compared to the quantum module Grover iteration in robust amplitude estimation (thus the former could be more NISQ friendly).
	Finally, we recommend two common pipelines suitable for real experiments to mitigate the noise effect based on our work and experiments. If the circuit size under investigation is small and the optimal parameters in the noiseless case can be obtained via numerical simulation, then neural + transformation retraining is strongly recommended since it costs no quantum computational resource overhead and offers sufficiently good performance for the ground state energy prediction. If the scale of the experiment is too large to simulate in silico, then we need to run the joint optimization from scratch for all parameters of different components (PQC, NN and transformation matrix). This pipeline essentially captures both the optimization process and the noisy joint retraining process simultaneously.
	
	\textbf{Conclusion:} In this work, we investigate the native QEM scheme for VQNHE and demonstrate that the adaptive retraining manifests excellent error-mitigating effects. We then analyze the QEM capacity 
	and present theoretical explanations for various scaling relations observed in experiments. In addition, we propose an enhancement add-on for VQNHE: the transformed Hamiltonian approach. Equipped with the parameterized gauge Hamiltonian, VQNHE++ shows even better expressive power and QEM capability. An interesting future direction is to apply VQNHE and the baked in error mitigation strategy shown in this work to more applications such as excited state searching problems or combinatorial optimization problems.
	
	~\newline
	\textbf{Acknowledgements:} This work is supported in part by the NSFC under Grant No. 11825404 (SXZ, ZQW, and HY), the MOSTC under Grants No. 2018YFA0305604 and No. 2021YFA1400100 (HY), the CAS Strategic Priority Research Program under Grant No. XDB28000000 (HY), and  Beijing Municipal Science and Technology Commission under Grant No. Z181100004218001 (HY).
	~\newline
	
	\textbf{Note Added:} After the completion of this work, we notice an interesting paper \cite{Shang} on related topics. This paper shares some similarities with the tri-optimization part in our work. While Ref. \cite{Shang} utilizes a bi-optimization setup of combining Heisenberg transformed Hamiltonian with variational quantum circuits, the present work employs a tri-optimization setup combining Heisenberg transformed Hamiltonian (potentially nonunitary), variational quantum circuit, and additionally neural networks in the middle with the help of VQNHE, which in general has larger expressive power.
	
%	\bibliographystyle{/Users/shixin/Cloud/refdatabase/apsreve.bst}
%	\bibliography{/Users/shixin/Cloud/refdatabase/lib3/library.bib}

	\clearpage
	\newpage
	
	\begin{widetext}
		\section*{Supplemental Materials}
		\renewcommand{\theequation}{S\arabic{equation}}
		\setcounter{equation}{0}
		\renewcommand{\thefigure}{S\arabic{figure}}
		\setcounter{figure}{0}
		\renewcommand{\thetable}{S\arabic{table}}
		\setcounter{table}{0}
		\renewcommand{\thesection}{S\arabic{section}}
		\setcounter{section}{0}
		
		\section{Single qubit example in detail: VQNHE, QEM and more}\label{smsec:onequbit}
		The motivations behind the calculation on the one-qubit system are: (1) the system is simple enough to be analytically traced and free from local minimum issues for retraining analysis since the number of trainable parameters is very limited, and (2) the system is still powerful enough for illustrating general features and providing insights on a general picture of the inherent QEM capacity of VQNHE.
		
		Specifically, we consider a system Hamiltonian defined on a single qubit as $\hat{H} = X+Z$ whose ground state is analytically given as $\vert \psi_0\rangle \propto (1-\sqrt{2}, 1)$ with the ground state energy $-\sqrt{2}$. We consider the depolarizing noise channel, where the ideal ground state density matrix $\rho_0 = \vert \psi_0\rangle \langle \psi_0\vert$ is transformed to $\rho = (1-p) \rho_0 +p I/2^n$. For a one-qubit system, the post-processing module $f$ has only one freedom $r=(f(1)-f(0))/(f(1)+f(0))$, note that the notation here is slightly different from the main text and the post-processing matrix is defined as
		\eq{\hat{f}=\begin{pmatrix} 1-r &0\\ 0&1+r\end{pmatrix}. }{arg2}
		
		The PQC in this example contains only one $\text{Ry}(\theta)$ rotation gate and thus the output wavefunction from the PQC without quantum noise is in the form of $(\cos(\theta), -\sin(\theta))$. The output density matrix in the presence of depolarizing quantum noise of strength $p$ is in the form of
		\eq{\rho = \begin{pmatrix}p/2 +(1-p)\cos^2\theta  & (p-1)\cos \theta\sin \theta \\(p-1)\cos \theta \sin \theta &p/2+(1-p)\sin ^2\theta )\end{pmatrix}.}{arg2}
		The final effective density matrix after VQNHE post-processing is $\rho_{\text {eff}} = \hat{f} \rho \hat{f}/\rm{Tr}(\hat{f}\rho\hat{f})$. With $p=0$ and $r=0$, we obtain the optimal parameters as $\theta_0 = -\arctan(1/(1-\sqrt{2}))\approx 1.178$ and the corresponding energy coincides with the exact value $-\sqrt{2}\approx -1.414$. With the noise $p>0$ turned on, the energy estimation with original weights is $E_{r_0, \theta_0}=-\sqrt{2} +\sqrt{2}p$. If only $\theta$ from the PQC is allowed to be retrained, the energy estimation cannot be improved. If only $r$ is allowed to be retrained, the optimal $r$ with noise and the corresponding energy estimation is given as $E_{r} =-\frac{\sqrt{2} (p+1)}{2 p+1}\approx  -2\sqrt{2} p^2+\sqrt{2} p -\sqrt{2}$ when $p\ll 1$, where the optimal new $r=\sqrt{2}p$. Therefore, the retraining energy gain is quadratic with the error strength $\delta E = E_{r}-E_{r_0}= -2\sqrt{2} p^2$.
		
		We further consider the case when both $r$ and $\theta$ can be further tuned on noisy hardware. In this case, the retrained energy estimation is $E_{r, \theta}\approx-\sqrt{2}+\frac{\sqrt{2}}{2} p$. Therefore, the QEM energy gain after joint retraining is $\delta E = E_{r, \theta} -E_{r_0, \theta_0} = -\frac{\sqrt{2}}{2}p $ which is better than retraining on the neural network only and shows linear scaling with the noise strength. In this case, the optimal parameters under noise is $\theta = \pi/4$ and $r=1/(1-p+\sqrt{2-2 p +p^2})\approx \left(1-\frac{1}{\sqrt{2}}\right)
		p+\sqrt{2}-1$. It is worth noting that the noise-aware optimal parameters here are not connected to the optimal parameter we have in the noise-free setup when $p\rightarrow 0$. This is the key for the linear scaling relation since we have proved any parameters adiabatically connected to the ideal case only contribute to quadratic scaling for the QEM capacity.
		
		In summary, the simple one-qubit example covers some of the main results in this work, including the scaling relation between the energy gain and the error strength as well as the reason behind such scaling relations.
		
		Finally, we investigate error-mitigated estimation on other observables instead of energy in the VQNHE setup. As we will show now, the retraining QEM scheme for VQNHE performs sub-optimally at predicting other observables and this is a general feature for variational wavefunction ansatz and not unique to our scheme. We focus on the observable $X$ and $Z$, the exact expectation is $\langle X\rangle=\langle Z \rangle = -1/\sqrt{2}$. In the presence of noise, the expectation under the noise-free optimal parameters is $\langle X \rangle = \frac{-1+p}{\sqrt{2}}$ and $\langle Z \rangle = \frac{-1+p}{\sqrt{2}}$. With the noise and neural only retrained parameters, we have the estimated expectation $\langle X\rangle =-\frac{(p-1) \left(2
			p^2-1\right)}{\sqrt{2} (2 p+1)}\approx -1/\sqrt{2}+3p/\sqrt{2}-2\sqrt{2}p^2 $ and $\langle Z\rangle = \frac{p (2 (p-1) p-3)-1}{\sqrt{2} (2
			p+1)}\approx -1/\sqrt{2}-p/\sqrt{2}$. Note the estimation of other observables' expectations does not get closer to the exact value after retraining. Namely, the retraining QEM strategy here is not suitable for error mitigation on other observables  than the Hamiltonian operator. In addition, with joint retraining, the estimated expectation becomes $\langle X\rangle =-\frac{2}{\left(\sqrt{p^2-2
				p+2}-p+1\right)
			\left(\frac{1}{\left(\sqrt{p^2-2
					p+2}-p+1\right)^2}+1\right)}\approx -1/\sqrt{2}+3p/(2\sqrt{2}) $ and $\langle Z\rangle = -\frac{2}{\left(\sqrt{p^2-2
				p+2}-p+1\right)
			\left(\frac{1}{\left(\sqrt{p^2-2
					p+2}-p+1\right)^2}+1\right)}\approx -1/\sqrt{2}-p/(2\sqrt{2})$. Again, the deviation of the estimation is worse, though slightly better than post-processing only retraining.
		
		\begin{table}[]
			\begin{tabular}{|cccc|}
				\hline
				&retraining energy gain & deviation of X & deviation of Z \\ \hline
				noise-free opt params &  0                     & $p/\sqrt{2}$                    & $p/\sqrt{2}$                                                \\
				neural only retraining    &  $-2\sqrt{2}p^2$                     &  $3p/\sqrt{2}-2\sqrt{2}p^2$                         & $-p/\sqrt{2}$                                             \\
				joint retraining      &  $-1/\sqrt{2}p$                     & $3p/(2\sqrt{2})$                          & $-p/(2\sqrt{2})$                                            \\ \hline
			\end{tabular}
			
			\caption{VQNHE-QEM results for one qubit system of Hamiltonian $\hat{H}=X+Z$ under depolarizing noise $p\ll 1$. 	\label{tab:onedp}}
		\end{table}
		
		Indeed, the variational wavefunction giving the minimal energy estimation is not guaranteed to give the most accurate expectations for other observables. Only when there is no noise and the variational expressive power is good enough, the variational wavefunction can approach the exact ground state wavefunction as closely as possible, and the estimation for other observables from such variational wavefunctions then become reliable. And this is not the case here, as the noise intrinsically forbids VQNHE to provide a sufficiently accurate approximation of the desired ground state under certain cases.  In other words, in the restricted Hilbert space accessible to VQNHE (limited by the noise, ansatz circuit, and choices of the neural network, etc.) for approximating the ground state, the state giving the lowest energy estimation does not necessarily coincide with the state giving the most accurate estimation on a given observable not commutable with the Hamiltonian operator. However, we stress that a worse estimation on another observable (other than the Hamiltonian) is not a unique side effect brought by our QEM scheme or quantum-neural hybrid state. Instead, this exact phenomenon exists for all variational algorithms in principle. The results for QEM energy gain and observable deviation before and after retraining in the presence of depolarizing noise are summarized in Table~\ref{tab:onedp}.
		
		We further consider another example: still the same system but the quantum circuit is subject to the influence of amplitude damping error channels instead.
		The Kraus operators for such error are defined as
		\eq{
			K_0 = \begin{pmatrix}1& 0\\
				0& \sqrt{1-\gamma}
			\end{pmatrix},~~K_1=\begin{pmatrix}0& \sqrt{\gamma}\\0&0\end{pmatrix}
			,}{}
		and the density operator in the presence of noise is $\rho = \sum_i K_i \rho_0 K_i^\dagger$.
		
		The energy estimation with the noise-free optimal parameters is $E_{r_0, \theta_0}=\frac{1}{2}
		\left(\left(2+\sqrt{2}\right)
		\gamma -\sqrt{2}
		\left(\sqrt{1-\gamma
		}+1\right)\right)\approx \left(1+3/
		(2\sqrt{2})\right) \gamma -\sqrt{2}$.
		As we can see, the energy deviation affected by the noise is still linear with $\gamma$ in the leading order, namely, $\gamma$ here characterizes the strength of the amplitude damping error, playing a similar role as $p$ in depolarizing channel. Meanwhile, the observable estimation with the ideal parameters are $\langle X\rangle = -\sqrt{1-\gamma}/\sqrt{2}\approx -1/\sqrt{2}+\gamma/(2\sqrt{2})$ and $\langle Z\rangle = -1/\sqrt{2}+(1+1/\sqrt{2})\gamma$.
		
		The energy gain by retraining only on the neural part (tuning $r$) is
		$$
		\delta E=-\frac{1}{2} \left(2+\sqrt{2}\right) \gamma
		+\frac{\sqrt{1-\gamma }}{\sqrt{2}}-\sqrt{\frac{1}{2
				\sqrt{2} \gamma +3 \gamma +1}+1}+\frac{1}{\sqrt{2}}=-\frac{\left(121+84 \sqrt{2}\right) \gamma ^2}{16
			\sqrt{2}}+O\left(\gamma ^3\right)
		$$, where the optimal $r\approx (3/2+7\sqrt{2}/8 )\gamma$ is adiabatically connect to $r=0$ when $\gamma\rightarrow 0$. The estimated expectation for other observables $X$ and $Z$ after retraining are $\langle X\rangle \approx \frac{3}{8} \left(4+3
		\sqrt{2}\right)
		\gamma-\frac{1}{\sqrt{2}}$ and $\langle Z\rangle \approx \frac{1}{8} \left(-4-3
		\sqrt{2}\right)
		\gamma-\frac{1}{\sqrt{2}}$ , respectively.
		
		Namely, the energy gain is still quadratic with the error strength. If the joint retraining is allowed, the energy can be fully recovered to the noiseless exact value. In this case, the estimated expectations for other observables are also exact. The gain part is thus linear with $\gamma$ as: $\delta E = -\left(1+\frac{3}{2 \sqrt{2}}\right) \gamma$. For the amplitude-damping channels, we find the scalings coincide perfectly with those for the case of depolarizing channel. Hence, we argue the universality of these QEM scaling relations. The results for QEM energy gain and observable deviation before and after retraining with the presence of amplitude damping noise are summarized in \ref{tab:one-ad}.

		\begin{table}[]
			
			\begin{tabular}{|cccc|}
				\hline
				&retraining energy gain & deviation of X & deviation of Z \\ \hline
				noise-free opt params &  0                     & $1/(2\sqrt{2})\gamma$                    & $(1+1/\sqrt{2})\gamma$                                                \\
				neural only retraining    &  $-\frac{\left(121+84 \sqrt{2}\right)}{16
					\sqrt{2}}\gamma^2$                     &  $\frac{3}{8} \left(4+3
				\sqrt{2}\right)
				\gamma$                         & $-\frac{1}{8} \left(4+3
				\sqrt{2}\right)
				\gamma$                                             \\
				joint retraining      &  $ -\left(1+\frac{3}{2 \sqrt{2}}\right) \gamma$                     & $0$                          & $0$                                            \\ \hline
			\end{tabular}
			
			\caption{VQNHE-QEM results for one qubit system of Hamiltonian $\hat{H}=X+Z$ under amplitude damping noise $\gamma\ll 1$. 		     \label{tab:one-ad}}
		\end{table}

		%	In other words, with variational wavefunction, only when the Hamiltonian expectation is very close to the exact ground state energy, the estimation on other observables becomes meaningful. Suppose the eigenstates are $\vert ii\rangle$, namely, the ground state is $\vert 0\rangle$. The observable we focus is $O$ which can be decomposed as $O = \sum_{i} O_{ii}\vert i\rangle \langle i\vert + \sum_{ij}O_{ij}\vert i\rangle \langle j\vert$.
		%	Meanwhile, the variational approximate ground state wavefunction is $\vert \psi\rangle = \sum_i c_i\vert i\rangle$. Therefore, the estimation on the expectation is $\langle \psi \vert O\vert \psi\rangle = \sum_i c_i^*c_i O_{ii}+\sum_{ij} c_i^* c_j O_{ij}$ while the exact value should be $O_{00}$. Thus, if the approximation is not good enough, the general expectation estimation deviates much from the exact ground state estimation.
		
		\section{Scaling results on biased neural retraining}\label{smsec:biased}
		We first recapitulate the setup for the biased neural retraining and explain why it is important to investigate the scaling between the biased retraining energy gain and the number of measurement shots required. For unbiased retraining, in each round of retraining optimization, we must take $k$ shots of measurement to ensure the sampling errors on quantum expectation and quantum gradient estimation are all under an acceptable threshold. In general, we need $t=O(10^3)$ optimization rounds to ensure the convergence of the energy estimations. Therefore, the total number of measurement shots required for unbiased training is $M=kt$ which may be very demanding on quantum resources in the NISQ era. Instead, for optimizations on the classical module such as the neural part or the transformation part in the transformed Hamiltonian approach, we can run the so-called biased retraining procedure, where only $M=k$ measurement shots are executed, and these $k$ bitstrings are utilized for each optimization round since the PQC does not change during the neural retraining. In other words, besides the $k$ measurement shots at the beginning of the biased retraining, all other workloads are done classically, rendering a more stable and light retraining procedure.
		
		However, the biased retraining workflow is {\bf biased}. Since $k$ measurement shots collected at the beginning can be biased due to the fine sampling errors, such bias may aggravate by the variational optimization. Namely, the retraining may overfit the biased quantum results and overestimate the QEM capacity. We must investigate the scaling between energy gain and the number of measurement shots carefully, so that we can differentiate the intrinsic QEM contribution from the biased overfitting contribution.
		
		To identify the intrinsic QEM capability from biased retraining, we first run the same PQC $N$ times ($N=819200$ for cases below unless specified) and save the corresponding sets of bitstrings. We then compute the energy gain averaged over retraining on randomly selected $N/100$ or $N/10$  bitstrings. Together with the energy retraining gain from the full set $N$ bitstrings, we find the average energy gain $\delta E$ scaling with the number of measurement shots required in the biased retraining.
		Since we compute the average energy gain, the measurement uncertainty for the energy estimation is suppressed by the order of $1/\sqrt{819200}$ which can be safely omitted.
		
		We use a one-dimensional five-site transverse field Ising model (TFIM) with open boundary conditions as VQNHE target Hamiltonian.  The Hamiltonian is given as $\hat{H}=\sum_{i=1}^{n-1} Z_iZ_{i+1}-\sum_{i=1}^n X_i$.
		The PQC is a layered ansatz [H, ZZ($\theta_1$), Rx($\theta_2$)].
		%The measurement uncertainty for retraining energy gain estimation is around $0.003$ with $N=819200$ measurement shots.
		We first apply the above scaling analysis to the ideal noise-free simulator. For $N/100, N/10, N$, the retraining energy gains are $-0.030, -0.0036, -0.0003$,  respectively. Since we expect no energy gain is possible when the retraining takes place in the noiseless setting, the non-vanishing values originate from overfitting the biased bitstring results. The scaling relation is  $\overline{\delta E} \propto 1/M$, where $M$ is the number of measurement shots the retraining is based on. In the infinite number of measurement shots limit $M\rightarrow \infty$, zero energy gain after retraining is recovered for the noiseless case.
		
		We further run the same biased retraining scaling analysis on IBM\_Santiago noisy simulator. The experiments without readout error mitigation give the energy gain $-0.0939$, $-0.0768$, $-0.0748$ for $N/100$, $N/10$, $N$ bitstrings. The scaling is perfectly described by $\overline{\delta E} = B + A/M$ where $B\approx -0.075$ is the intrinsic QEM part contributing to the energy gain from neural retraining and $A/M$ scaling part gives the overfitting artifact. The same scaling relation applies for noisy simulator results with readout error mitigation and results directly collected from IBM\_Santiago hardware.
		
		For IBM\_Santiago simulator, after enforcing the readout error mitigation, the raw results are improved and the retraining gain is lower:  the gains are
		$-0.0366$, $-0.0116$, $-0.009$ for $8192$, $81920$, $819200$  bitstring results based retraining. The intrinsic QEM offset is around $B\approx -0.009$, less than the case without readout error mitigation in magnitude ($-0.075$).
		
		Besides simulation, we carry out the scaling analysis experiments on IBM\_Santiago hardware, where $8192*50$ sets of bitstrings are collected in total. The retrained energy gains without the readout error mitigation are $-0.135$, $-0.113$, $-0.112$ for results based on $8192$, $81920$ and $8192*50$ bitstrings, respectively. Again, the effect of retraining on the energy gain still follows the $\overline{\delta E} = B+A/M$ scaling with intrinsic QEM capacity$B\approx -0.111$.
		
		To further investigate the QEM effect of the retrained VQNHE in the presence of quantum noise, we utilize depolarizing error model, where each two-qubit gate is followed by applying a depolarizing noise of strength $p$ on each qubit. For a given p, say $p=0.02$, we observe the same overfitting scaling relation between the biased energy gain and the number of measurement shots based. For $M=8192,81920, 819200$, we have the energy gain as $-0.0694, -0.0482$ and $-0.0464$, respectively, where the scaling relation is approximated by $\overline{\delta E} \approx -0.046 - A/M$ again.
		
		Finally, we comment that the scaling here is for the average of $\delta E$, where the average is taken over different groups composed of $M$ measurement shots. For individual cases, the energy gain retrained on $M$ measurement shots is a random variable, with the mean value scale as $B+A/M$ as indicated before and the standard deviation is in the order ${1}/{\sqrt{M}}$, which is the typical behavior due to the finite sampling errors.
		
		\section{Energy landscape for joint retraining}\label{smsec:joint}
		
		See Fig.~\ref{fig:landscape} on why perturbation understanding fails in the joint retraining scenario and how the introduction of noise breaks the symmetry between different noiseless optimal solutions.
		
		\begin{figure}[t]\centering
			\includegraphics[width=0.65\textwidth]{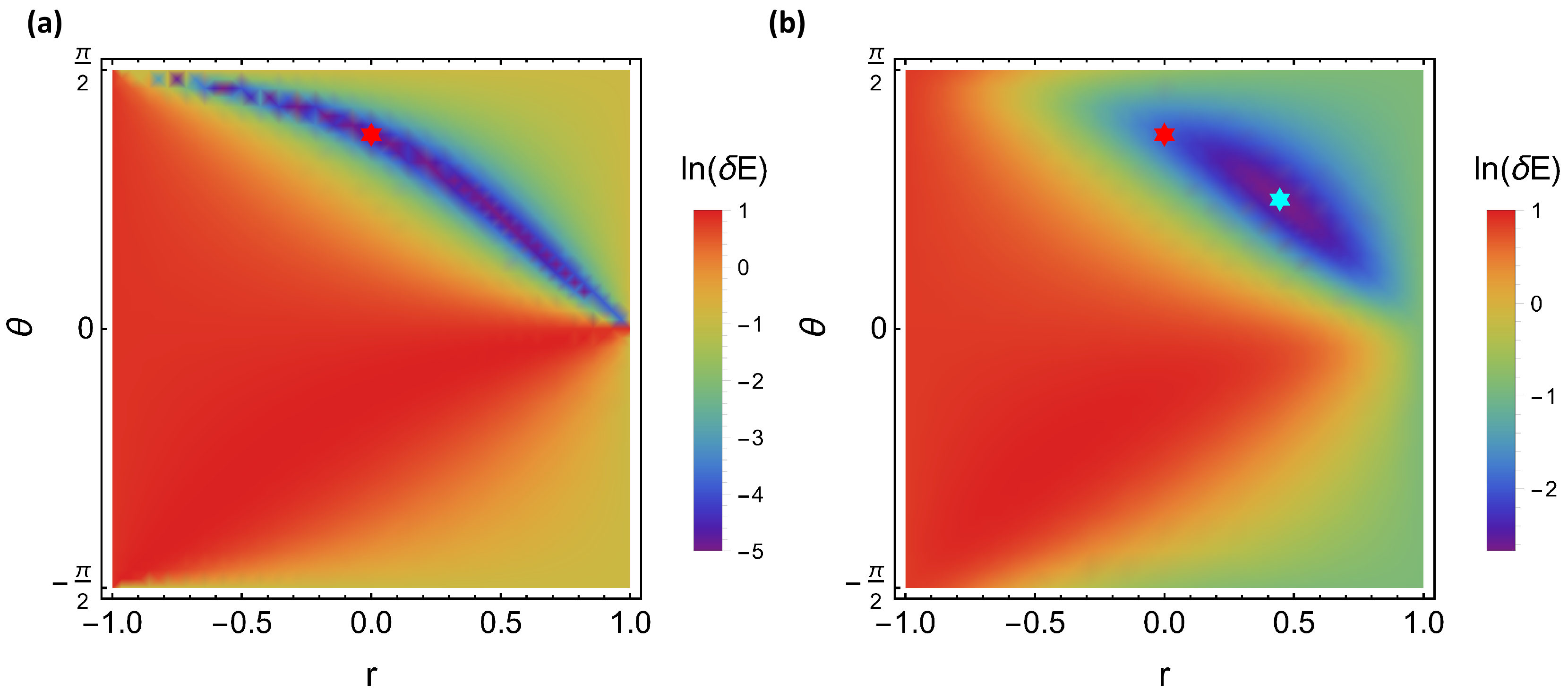}
			\caption{Energy landscape with varying $r$ and $\theta$ for one-qubit system $\hat{H}=X+Z$. The energy is presented in log scale $\ln(E-E_0+e^{-5})$, where $E_0$ is the exact ground state energy and $E$ is the energy estimation given circuit parameter $\theta$ and neural parameter $r$. (a) Noiseless case: the red star indicates the ideal solution with $r=1$. All points in the deep blue region of value $-5$ are possible optimal solutions. (b) Depolarizing noise case with $p=0.1$: ideal solutions respond differently with the noise on, and the most noise resilient point is indicated by the cyan star which is not adiabatically connected to the original red star. This non-perturbative nature is the origin of linear scaling for QEM capacity.   }
			\label{fig:landscape}
		\end{figure}
	
	\section{Scaling of retraining energy and the noise strength}\label{smsec:logscale}
	
	In this section, we fit the numerical data for retraining energy gain $\delta E$ and the corresponding overall depolarizing noise strength $p_{eff}$ in log-log scale. In the main text, we fit the curve with fixed power $1$ and $2$ in each case by only identifying the optimal prefactor. In this part, we fit in log scale without prior knowledge of the power, i.e. we use the linear fit with both the prefactor and the scaling power as unknown parameters. The obtained results are consistent with our conclusion and demonstrate our picture. See Fig. \ref{fig:sm_log_scaling} for the scaling relation in log-log axis with $\ln (-\delta E)$ and $\ln p_{eff}$. The fitting relation are $\delta E = -10.07\times p_{eff}^{1.98}$ and $\delta E = -1.91\times p_{eff}^{0.93}$ for neural retraining and joint retraining cases, respectively.
	
	\begin{figure}[t]\centering
		\includegraphics[width=0.58\textwidth]{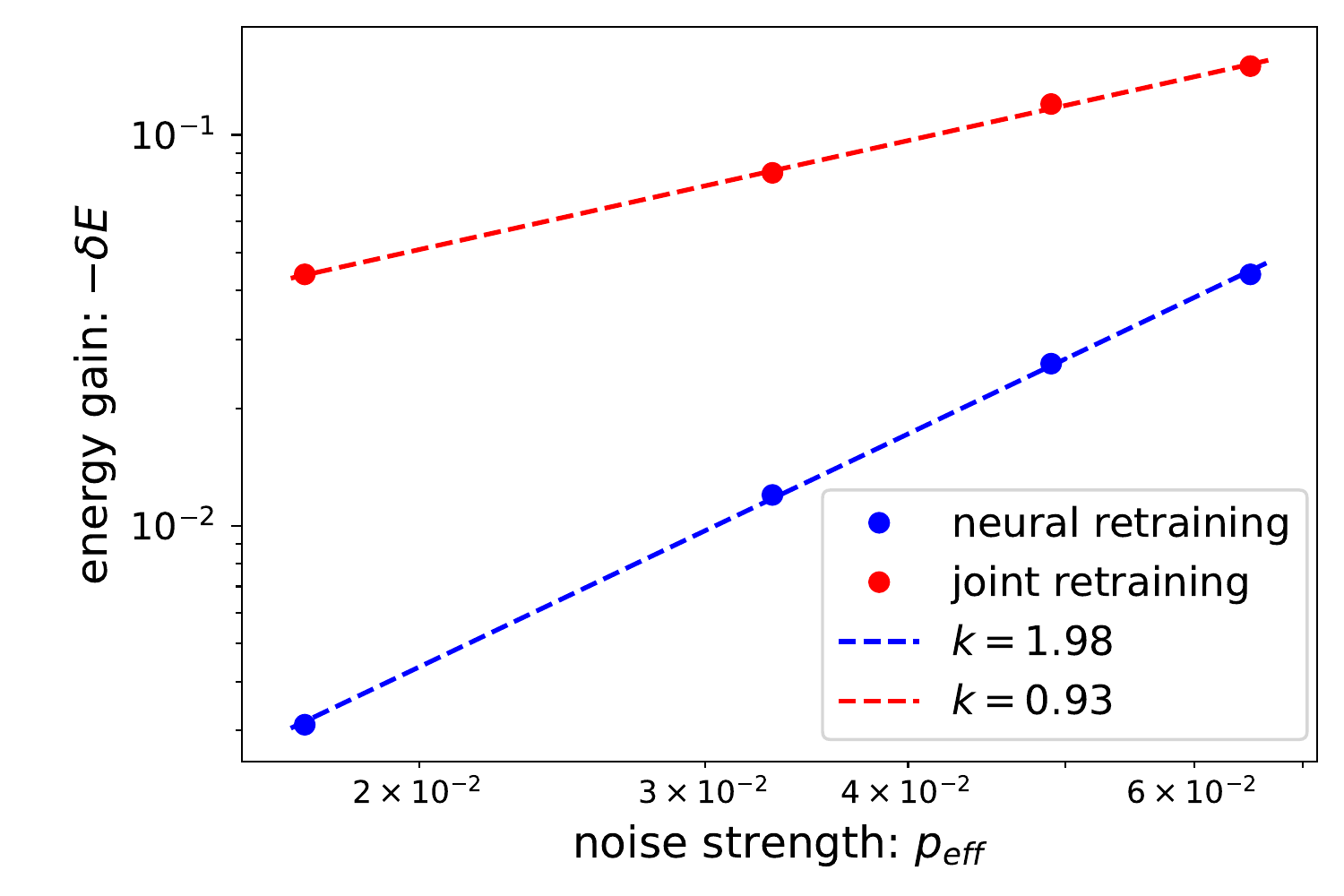}
		\caption{Scaling relation between retraining energy and the noise strength in log-log scale. The system is five-qubit TFIM and the noise model is depolarizing after each two-qubit gate. The powers given by the fit are very close to $1$ and $2$, respectively, consistent with our theoretical picture.}
		\label{fig:sm_log_scaling}
	\end{figure}
		
		\section{Layered quantum ansatz notation}\label{smsec:notation}
		
		We use a list notation for layered circuit ansatz as [A, B, C...] where each term represents a quantum layer. The layers we utilized in this work include (suppose $n$ qubits for the circuit):
		\begin{itemize}
			\item H for Hadamard layer: $\prod_{i=1}^n H_i$ where $H_i$ is Hadamard gate on the $i$-th qubit.
			\item ZZ($\theta$) for parameterized ZZ layer: $\prod_{i=1}^{n-1} e^{i\theta_i Z_iZ_{i+1}}$ where $\theta_i$ is trainable weights and $Z_i$ is the Pauli Z gate on the $i$-th qubit. Similar rules apply to XX and YY layer.
			\item Rz($\theta$) for parameterized rotation layer around z axis: $\prod_{i=1}^n e^{i\theta_i Z_i}$. Similar rules apply to Rx and Ry layer.
			\item SWAP($\theta$) for parameterized SWAP layer: $\prod_{i=1}^{n-1} e^{i\theta_i \text{SWAP}_{i,i+1}}$, where SWAP=$\begin{pmatrix}1&0&0&0\\0&0&1&0\\0&1&0&0\\0&0&0&1\end{pmatrix}$.
		\end{itemize}
		
		For example, the parameterized circuit ansatz in the form [H, ZZ($\theta$), Rx($\theta'$)] can be expressed as:
		\eq{U=\prod_{i=1}^n e^{i\theta'_i X_i}\prod_{i=1}^{n-1} e^{i\theta_i Z_iZ_{i+1}}\prod_{i=1}^n H_i.}{}
		
		\section{Gauge transformation ansatz choice}\label{smsec:gauge}
		The parameterized gauge transformation has one requirement for the scalability of VQNHE++: the Hamiltonian after the transformation $H'=W^\dagger HW$ has to contain only polynomial terms of Pauli string so that the expectation can be efficiently measured on a quantum computer. The commonly utilized scalable parameterized transformation is single-qubit rotation on each qubit as $W=\prod_i e^{i\tau_iP_i}$, where $P_i$ is the local Pauli operator on site $i$.
		
		For a spin Hamiltonian with maximal Pauli string length $k$, such a local gauge transformation can induce new terms of Pauli string from a Pauli string $\prod_{i\in S_k} Q_i$ ($Q_i$ is local Pauli operator while $S_k$ is the index set of size $k$)
		
		\eq{\prod_{i\in S_k} e^{-i\tau P_i} (\prod_{i\in S_k} Q_i)  \prod_{i\in S_k} e^{i\tau P_i} = \prod_{i\in S_k} (\cos \tau_i + i \sin\tau_i P_i)Q_i (\cos \tau_i - i \sin\tau_i P_i)}{}
		
		Namely, if we have $m$ terms in the original Hamiltonian, the transformed Hamiltonian with local gauge transformation will have at most $4^k m$ terms. Therefore, for spin models with generally $k\sim O(1), m\sim O(N)$, the resulting Hamiltonian still contains polynomial terms of Pauli string as $O(N)$ For molecule Hamiltonian, with Bravyi-Kitaev mappings, we have $k\sim O(\ln N) $ and $m\sim O(N^4)$. The transformed Hamiltonian contains  $O(N^5)$ Pauli string terms, which is still in polynomial scaling.
		
		The local gauge transformation ansatz can be extended to the transformation unitary defined as a shallow circuit. The perspective is that, for each Pauli string, the transformed Pauli string must be in the casual light cone via the shallow circuit transformation. As long as the transformation ansatz is shallow, the support of the transformed Pauli string is as small as $k$, and at most $4^k$ terms of Pauli string can emerge in the transformed Hamiltonian. For example, for 2-qubit Pauli strings, with the transformation ansatz as one layer of even-odd brick-wall two-qubit, gates, the light cone can cover $k=6$ qubits at most (two layers of brick-wall two-qubit gate gives $k=10$). Therefore, we can maintain the polynomial scaling for the number of Pauli string terms in the transformed Hamiltonian, though the constant factor of the scaling increases fast with the ansatz circuit depth. 
		
		It is also worth noting that the Clifford circuit is also one type of scalable transformation ansatz, since the ansatz can map one Pauli string to another Pauli string by definition. In this case, the ansatz is parameterized by some discrete variables that control the structure of Clifford ansatz instead of the continuous variable as introduced above. So we can utilize gradient-free optimizers such as Bayesian optimization to optimize the gauge transformation in this case.
		
		\section{Nonunitary version of transformed Hamiltonian approach}\label{smsec:nonunitary}
		
		For the transformed Hamiltonian approach, the transformation can also be nonunitary. For nonunitary $\hat{W}$, we regard it as the nonunitary operation on the VQNHE output state as $\vert \psi_W\rangle = \frac{\hat{W}\vert \psi_f\rangle}{\vert \hat{W}\vert \psi_f\rangle\vert}$. Therefore, the final energy estimation is given by
		\eq{\langle \hat{H}\rangle=\frac{\langle 0^n\vert U^\dagger \hat{f}^\dagger\hat{W}^\dagger \hat{H}\hat{W} \hat{f}U\vert 0^n \rangle}{\langle 0^n\vert U^\dagger \hat{f}^\dagger\hat{W}^\dagger \hat{W} \hat{f}U\vert 0^n \rangle}.}{}
		
		In experimental implementations, the nonunitary operation of neural module $\hat{f}$ is simulated by the VQNHE measurement scheme and the nonunitary operation of transformation module $\hat{W}$ is simulated by transforming the Hamiltonian classically. The effective overall nonunitary quantum channel $\hat{W}\hat{f}$ greatly enhances the expressive power and quantum noise resilience of plain VQE where only the quantum circuit $U$ can be tuned. Typical examples for nonunitary transformation are also single-qubit rotations $\rm{exp}(i\tau \hat{P})$, but $\tau$ can take complex value this time.
		
		Note that we have actually incorporated the nonunitary property of the transformation in the TFIM example in the main text. However, the results after optimization all give zero imaginary part in $\tau$, indicating nonunitary character has no further gain in TFIM + single-qubit rotation transformation case. As we will see, this is not true for Heisenberg model simulation where nonunitary part of the transformation plays an important role.

		\section{Transformed Hamiltonian VQNHE results on Heisenberg model}\label{smsec:heisenberg}
		In this section, we report tri-optimization results on one-dimensional six-sites isotropic Heisenberg model with an open boundary condition, whose Hamiltonian is given by: $\hat{H}=\sum_{i=1}^{n-1}(X_iX_{i+1}+Y_{i}Y_{i+1}+Z_iZ_{i+1})$. The ground state energy is $-9.9743$. Since the system has SU(2) symmetry, we pick the circuit ansatz which also respects such symmetry and thus conserves the total spin $J_{\text{tot}}^2$. In this case, we adopt a symmetry-preserving PQC layout: [SWAP($\theta_1$), SWAP($\theta_2$)]. Also, we choose a transformation that keeps the symmetry instead of a single-qubit rotation layer. To keep the polynomial overhead for the transformed Hamiltonian, we utilize ``half-layer" of parameterized SWAP as the transformation. Specifically, the parameterized transformation we utilized is:
		\eq{\hat{W}_\tau=\prod_{i=1,3,5} e^{i\tau_i \text{SWAP}_{i,i+1}},}{arg2}
		where $\tau$ can take complex values as explained in the above section. Such transformation has geometrically compatible gates which are classically tractable and leads to a transformed Hamiltonian containing a polynomial number of local terms.
		
		Apart from the new choice on the parameterized transformation, we also allow the post-processing neural model to output complex values, which further enhances the power of the end-to-end setup. We still consider the same type of quantum noise: depolarizing channel attached after each two-qubit gate. And the optimized results with different retraining strategies are shown in Fig.~\ref{fig:hei}.

		Fig.~\ref{fig:hei} conveys a few very important messages. Firstly, we again validate that pure retraining on the PQC gains very little, which supports our conclusion that the noise robustness is unique to VQNHE instead of VQE (corresponding to q retraining). Secondly, the half layered parameterized SWAP transformation is very powerful as it is implemented classically without noise while keeping the quantum state in the correct symmetry factor. The mitigated energy is at least $E=-9$ since even for the fully mixed state $\rho=I/2^n$, the transformation, as a quantum channel effectively, can project the system to an averaged energy of $-9$. Lastly, we again observe that while triple retraining gives the best error mitigation capacity, the pure classical optimization with n+t is still good enough and even outperforms q+n retraining. Therefore, we can run biased retraining on n+t very fast and obtain much more reliable energy estimations without using additional quantum resources.
		
		To show the universal QEM capacity for our method, we also show the triple optimization setup for Heisenberg model VQE with pure dephasing quantum error after each two-qubit gate. The results are shown in Fig.~\ref{fig:heisenbergdephasing}.
		
				\begin{figure}[t]\centering
			\includegraphics[width=0.6\textwidth]{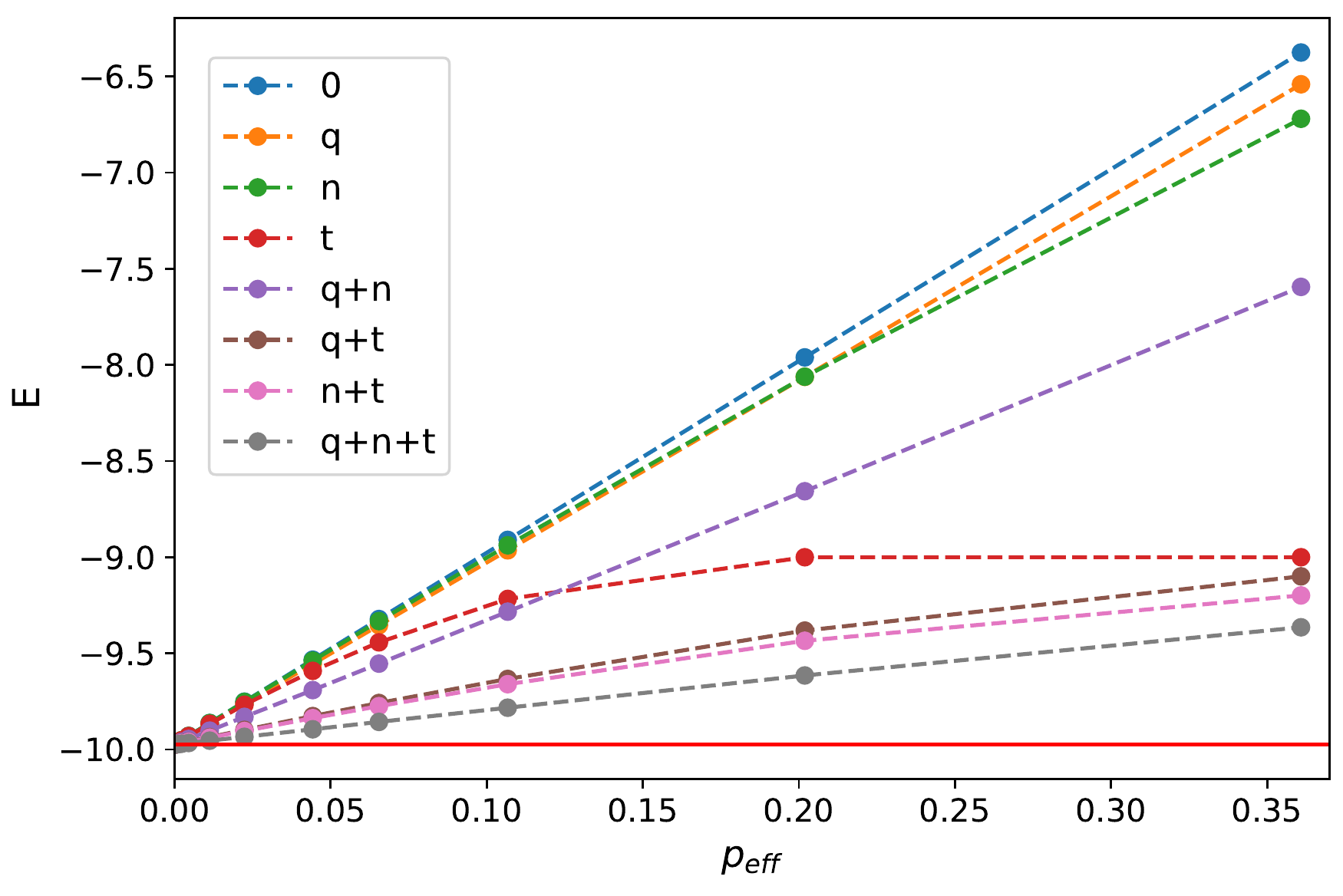}
			\caption{VQNHE++ for 1D Heisenberg model with overall pure dephasing noise $p_{\text {eff}}$. 0 indicates the energy estimation with noiseless optimal weights (no retraining). q, n, t is for retraining on the PQC, neural network and parameterized transformations, respectively. The solid line is the exact ground state energy for the simulated system.}
			\label{fig:heisenbergdephasing}
		\end{figure}
		
		\section{Hyperparameter settings}\label{smsec:hyper}
		Most numerical simulations in this work are conducted using the tensor network based differentiable quantum simulator: TensorCiruit (\url{https://github.com/tencent-quantum-lab/tensorcircuit}). We use the density matrix simulator {\sf DMCircuit} to exactly characterize the behavior for the quantum noise which takes twice the number of qubits as the conventional state simulators.
		
		The neural model for post-processing in VQNHE is a simple fully connected neural network with the output range $[1/e, e]$. The detailed choice of the model is irrelevant since the model easily has full expressive power for the small system size we simulated in this work. For the complex-valued post-processing module we tested in tri-optimization setup on Heisenberg model, we simply utilized a $2^n$-dimensional complex-valued variational vector $f$ as the post-processing model. Note that we don't impose the output range restriction in this complex valued case, which may lead to better error mitigation results. In the case when $n$ is large, we believe neural networks commonly used in variational Monte Carlo scenarios are sufficient to use in VQNHE setup. The representative neural network structures include restricted Boltzmann machine, recurrent neural network and transformers. Whether there are differences in terms of  error mitigation capacity for different neural network structures is an interesting future direction.
		
		In the tri-optimization setup, the three sets of parameters are updated simultaneously. We apply three Adam optimizers with learning rates $0.005, 0.01$ and $0.003$ on quantum module, neural module and transformation module, respectively. The convergence of the optimization usually takes thousands of epochs from random initialization.
		
		The initialization on both the PQC and the transformation parameters are drawn from Gaussian distribution with zero mean and small standard deviations, say $0.1$. The initialization near zero ensures that the initial effect of these modules behaves similarly to identity operations which is helpful for a stable training process later.
		
	\end{widetext}
	
\end{document}